\documentclass{article}
\usepackage{graphicx}
\usepackage{cite}
\usepackage{epsfig}
\usepackage{amsfonts}
 \usepackage{amssymb}
\usepackage{amsmath}
\usepackage{slashed}

\usepackage{hyperref}
\usepackage{graphicx}       
\usepackage{dcolumn}        
\usepackage{bm}            		 
\usepackage{amssymb}
\usepackage{amstext}
\usepackage{tensor}
\usepackage{color} 
\usepackage{transparent}

\date{\today}

\newcommand{\insertplot}[5]{\begin{figure}
 \hfill\hbox to 0.05in{\vbox to #5in{\vfill
 \inputplot{#1}{#4}{#5}}\hfill}
 \hfill\vspace{-.1in}
 \caption{#2}\label{#3}
 \end{figure}}
 \newcommand{\inputplot}[3]{
 \special{ps: plotfile #1}
}
\newcounter{fig}

\newcommand{\ee}{\end{equation}}
\newcommand{\eea}{\end{eqnarray}}
\newcommand{\bea}{\begin{eqnarray}}

\newcommand{\beq}{\begin{equation}}
\newcommand{\eeq}{\end{equation}}

\newcommand{\ze}{\kern 0.05em}

\begin{document}
\begin{center}

{\LARGE \bf Einstein-Maxwell-scalar black holes: \vspace{0.2cm} \\ the hot, the cold and the bald}
\vspace{0.8cm}
\\
{{\bf Jose Luis Bl\'azquez-Salcedo$^{\Diamond}$,
Carlos A. R. Herdeiro$^{\ddagger}$,  \\
Jutta Kunz$^{\Diamond}$,
Alexandre M. Pombo$^{\ddagger}$ and 
Eugen Radu$^{\ddagger}$ 
}
\vspace{0.3cm}
\\
$^{\Diamond}${\small Institut f\"ur  Physik, Universit\"at Oldenburg, Postfach 2503,
D-26111 Oldenburg, Germany}
\vspace{0.3cm}
\\
$^{\ddagger }${\small Departamento de Matem\'atica da Universidade de Aveiro and } \\ {\small  Centre for Research and Development  in Mathematics and Applications (CIDMA),} \\ {\small    Campus de Santiago, 3810-183 Aveiro, Portugal}
}
\vspace{0.3cm}
\end{center}

\date{September 2019}

\begin{abstract}   
The phenomenon of spontaneous scalarisation of charged black holes (BHs) has  recently motivated studies of various Einstein-Maxwell-scalar models. Within these models, different classes of BH solutions are possible, depending on the non-minimal coupling function $f(\phi)$, between the scalar field and the Maxwell invariant. Here we consider the class wherein  both the (bald) electrovacuum Reissner-Nordstr\"om (RN) BH and new scalarised BHs co-exist, \textit{and} the former are never unstable against scalar perturbations. In particular we examine the model, within this subclass, with a quartic coupling function: $f(\Phi) = 1+\alpha \Phi ^4$. The domain of existence of the scalarised BHs, for fixed $\alpha$, is composed of two branches. The first branch (\textit{cold} scalarised BHs) is continuously connected to the extremal RN BH. The second branch (\textit{hot} scalarised BHs) connects to the first one at the minimum value of the charge to mass ratio and it includes overcharged BHs. We then assess the perturbative stability of the scalarised solutions, focusing on spherical perturbations. On the one hand, cold scalarised BHs are shown to be unstable by explicitly computing growing modes. The instability is quenched at both endpoints of the first branch. On the other hand, hot scalarised BHs are shown to be stable  by using the S-deformation method. Thus, in the spherical sector this model possesses two stable BH local ground states (RN and hot scalarised).   We point out that the branch structure of BHs in this model parallels the one of BHs in five dimensional vacuum gravity, with [Myer-Perry BHs, fat rings, thin rings] playing the role of [RN, cold scalarised, hot scalarised] BHs. 
\end{abstract}

\tableofcontents

\newpage

%
 \section{Introduction}\label{S1}
%
Einstein-Maxwell-scalar (EMS) models described by the action
\begin{equation}\label{E21}
	 \mathcal{S}=\int d^4 x \sqrt{-g} \Big[R-2\partial _\mu \Phi \partial ^\mu \Phi -f (\Phi) F_{\mu \nu} F^{\mu \nu} \Big]\ ,
\end{equation}
where $R,F_{\mu\nu},\Phi$ are, respectively, the Ricci scalar,  Maxwell 2-form and a real scalar field, have been shown to allow the phenomenon of spontaneous scalarisation of asymptotically flat, charged black holes (BHs)~\cite{Herdeiro:2018wub,fernandes2019spontaneous}. In the models where this phenomenon occurs,  the standard electrovacuum Reissner-Nordstr\"om (RN) BH is an equilibrium \textit{scalar-free} (or \textit{bald}) solution. For sufficiently large BH charge to mass ratio, however, the RN BH becomes unstable against scalar perturbations. This perturbative instability is stalled by non-linear effects. The process can be followed dynamically and a new BH configuration forms as the endpoint of the evolution, possessing a non-trivial scalar field profile. These new BH configurations can also be obtained as equilibrium solutions of the field equations. Under certain conditions, these \textit{scalarised} (or \textit{hairy}) BHs are the preferred vacuum.

Both the possible existence of a RN solution of~\eqref{E21} and the possible instability of this solution against scalar perturbations, depend on the choice of the non-minimal coupling function $f(\phi)$. A classification of models according to the sort of BH solutions allowed by the choice of $f(\phi)$ was suggested in~\cite{Astefanesei:2019pfq}. 
Class I model BHs have always a nonvanishing scalar field; in particular, the RN BH is not a solution. 
A representative example of this class is the Einstein-Maxwell-dilaton model.
Amongst the class of models that allow both the scalar-free RN solution and new scalarised BH solutions, subclasses IIA and IIB models were distinguished in~\cite{Astefanesei:2019pfq} to account for two different possibilities: models in which the RN BH is (IIA)  or is not (IIB) unstable against scalar perturbations. Studies up to now have focused on models of subclass  IIA, see $e.g.$~\cite{Herdeiro:2018wub,Myung:2018vug,Boskovic:2018lkj,Myung:2018jvi,fernandes2019spontaneous,Brihaye:2019kvj,Herdeiro:2019oqp,
Myung:2019oua,Astefanesei:2019pfq,Konoplya:2019goy,Fernandes:2019kmh,Herdeiro:2019tmb,Zou:2019bpt,Brihaye:2019gla}. The purpose of this work is to consider models of  subclass IIB.

 Subclass IIB models raise immediately two questions. Firstly, since the RN BH is not unstable against scalar perturbations, there is no zero mode solution. Thus, it is unclear if the scalarised solutions in this subclass bifurcate from RN at all, or if they are completely disconnected. Secondly, are scalarised solutions stable or unstable? Since the RN solutions in this subclass are stable against scalar perturbations, stable scalarised solutions would imply the existence of two stable, possibly disconnected, local ground states.

In this paper we will consider an illustrative case of subclass IIB models, by taking a quartic coupling function. This will allow us to clarify the first question above: indeed the scalarised BHs 
 are connected with the
RN solution, but this occurs only at extremality. 
This is qualitatively different from subclass IIA models, wherein bifurcation occurs for non-extremal BHs that admit a zero mode of the scalar field. Moreover, in subclass IIA models, the scalarised BHs have no extremal limit~\cite{Herdeiro:2018wub,fernandes2019spontaneous}, unless they carry also a magnetic charge~\cite{Astefanesei:2019pfq}. The second question is more intriguing. We shall provide a partial answer, by showing that one of the branches of scalarised solutions (corresponding to the \textit{hot} scalarised BHs) is stable against spherical perturbations. Thus, at least in the spherical sector, this model allows for two stable local ground states. Again, this is qualitatively different from subclass IIA models, wherein scalarised solutions are perturbatively stable, but RN BHs are unstable, even restricting to the spherical sector.

This paper is organised as follows. General comments on EMS models are given in Section~\ref{S2}, including a description of the equations of motion, boundary conditions and universal relations. The analysis of the chosen illustrative example of subclass IIB models is made in Section~\ref{S3}. This includes obtaining the domain of existence for BH solutions, the description of the cold and hot branches and 
their connection with RN at extremality. In Section~\ref{S4} we perform the stability analysis of the scalarised solutions. For the cold branch, unstable quasi-normal modes are explicitly shown to exist. For the hot branch, a proof of stability against spherical perturbations is given, using the S-deformation method~\cite{Kimura:2017uor,Kimura:2018eiv,Kimura:2018whv}. Final remarks, including a curious analogy between the space of solutions of this model and that of five dimensional vacuum gravity, are presented in Section~\ref{S5}.

%
 \section{The EMS model}\label{S2}
%
The action of the family of EMS models we wish to consider is given by~\eqref{E21}. We are interested in spherical BH solutions. 
  For the metric, we consider the spherically symmetric ansatz parameterised as 
	\begin{equation}\label{E22}
	 ds^2 = - N(r)\ze e^{-2  \delta (r)}
	dt^2+\frac{dr^2}{N(r)}+r^2 \big(d\theta ^2 +\sin ^2 \theta  d \phi ^2\big)\ ,~~{\rm with
	}~~N(r)\equiv 1-\frac{2m(r) }{r},
	\end{equation}
	where $m(r)$ is the Misner-Sharp mass function~\cite{misner1964relativistic}. 
	Spherical symmetry requires the scalar field to depend only on the radial coordinate, $\Phi=\Phi\ze (r)\ze $. The Maxwell tensor is $F_{\mu \nu}=\partial _\mu A_\nu -\partial _\nu A_\mu $, where the $4-$vector potential has only a non-trivial spherical electrostatic potential, $A_\mu =\big( A_t(r),0,0,0\big)$. A magnetic charge would also be compatible with spherical symmetry, but that shall not be considered here -  see~\cite{Astefanesei:2019pfq} for magnetically charged BHs in this context. 
	
In spherical symmetry, the BH solutions of this model can be studied using the effective Lagrangian
	\begin{equation}\label{E25}
 \mathcal{L}_{\rm eff} =e^{-\delta (r)}m'(r)
-\frac{1}{2}r^2 e^{-\delta(r)}N(r)\Phi ^{'2}(r)
+\frac{1}{2}f(\Phi) e^{\delta (r)}r^2 A_t^{' 2 }(r) ~.
	\end{equation}
Here and below a prime denotes a derivative with respect to the radial coordinate, $e.g$, $\Phi'\equiv d \Phi /dr$; also, below,  $\dot{f} (\Phi)\equiv d f (\Phi )  /d\Phi $. For ease of notation we shall often omit the radial dependences.

From the Lagrangian \eqref{E25}, one obtains the field equations: 
	\begin{eqnarray}
\label{E29}
&&
m'=\frac{r^2N \Phi ^{' \ze\ze 2}}{2} +\frac{Q^2}{2r^2 f(\Phi)} \ , \qquad \delta'+r \Phi ^{' \ze\ze 2}=0 \ , \qquad 
 A_t'  = -\frac{Q e^{-\delta}}{f (\Phi) r^2} \ ,
\\
&&
\Phi ^{''} (r)+\frac{1+N}{rN}\Phi ^{'} 
-\frac{Q^2}{r^3N f (\Phi)  }
\left(
\Phi ^{'}
-\frac{\dot{f} (\Phi)}{2r  f (\Phi)}
\right)=0~.
\end{eqnarray}	
	The electric potential equation \eqref{E29} yields a first integral which was used to simplify the remaining equations. The constant of integration is interpreted as the electric charge $Q$.
The solutions of these equations will be obtained numerically. As such, the behaviour of the different functions near the boundaries of the integration domain (horizon and spatial infinity) need to be considered.

	At the horizon, the field equations can be approximated by a power series expansion in $r-r_H$, that depends only on the horizon values of some of the functions, denoted by a subscript $H$ ($e.g.$, $\sigma _H $, $\Phi _H$) and horizon radius $r_H$,
			\begin{eqnarray}
&& m(r) = \frac{r_H}{2}+ \frac{Q^2}{2r_H ^2 f(\Phi_H)} (r-r_H) +\cdots\ \ , \qquad
			\delta (r)  = \delta _H -\Phi_1^2 r_H (r-r_H)+\cdots\ ,
			\\
&& 
\nonumber
\Phi ( r) =  \Phi _H + \Phi_1 (r-r_H)+\cdots\ , \qquad \qquad \ \ \ \ \ 
			A_t(r) = -\frac{e^{-\delta _H }Q}{r_H ^2 f (\Phi_H)} (r-r_H)+\cdots\ ,
\end{eqnarray}
where
\begin{equation}
			 \Phi_1=\frac{Q^2\dot{f} (\Phi_H)}{2 r_H f(\Phi_H) \big[Q^2-r_H^2 {f} (\Phi_H)\big]}~.
\end{equation}			
	At spatial infinity, on the other hand, the field equations can be approximated by a power series expansion in $1/r$,
	\begin{eqnarray}
	&&
	m(r) =M- \frac{Q^2+Q_s^2}{2r}+\cdots\ ,\qquad \qquad \qquad   
	\delta (r) \approx  \frac{Q_s^2}{2r^2}+\cdots\ ,
	\\
	&&
	\nonumber
	\Phi (r) = \frac{Q_s}{r}+\frac{MQ_s}{r^2}+\cdots\ ,\qquad \qquad \qquad \quad A_t(r) =\mathcal{V}-\frac{Q}{r}+\cdots \ ,
	\end{eqnarray}
{where we have assumed $f(0)=1$.}
Here, $M,Q,Q_s$ denote, respectively, the ADM mass, BH electric charge and the scalar ``charge". The latter, albeit not a gauge charge, determines the monopolar term, $i.e.$ the leading decay term of the scalar field at infinity. $\mathcal{V}$ is the electrostatic potential at infinity.

Integrating the field equations with these asymptotic behaviours one finds a family of BH solutions. The solutions can be physically characterised by the following dimensionless quantities: the charge to mass ratio, $q$, the reduced horizon area, $a_H$, and the reduced horizon temperature, $t_H$, 
\begin{equation}
q\equiv \frac{Q}{M} \ , \qquad  a_H\equiv \frac{A_H}{16\pi M ^2} = \frac{ r_H ^2}{4 M ^2}\ ,\qquad t_H\equiv 8\pi M T_H= 2MN'(r_H) e^{-\delta (r_H)}\  \ ,
\end{equation}
where $A_H,T_H$ are the area and temperature of the BH.

To assess the regularity of the solutions, one may consider curvature invariants, such as the Ricci scalar $R$ and the Kretschmann scalar ($K\equiv R_{\mu \nu \delta \lambda} R^{\mu \nu \delta \lambda}$). For our solutions, they read 
\begin{equation}
	 R=\frac{N'}{r}(3r\delta '-4)+\frac{2}{r^2}\Big\{ 1+N \big[ r^2 \delta '' -(1-r \delta ')^2\big]\Big\} -N''\ , 
\end{equation}
and 
\begin{equation}
K=\frac{4}{r^4}(1-N)^2+\frac{2}{r^2}\left[N^{'2}+(N'-2N\delta')^2\right]+\left[N''-3\delta' N' +2N(\delta^{'2}-\delta'')\right]^2\ .
\end{equation}
\bigskip
	To test our numerics we use three different checks. First, the following virial identity 
	\begin{equation}
	 \int _{r_H} ^{\infty} dr \left\{e^{-\delta }r^2\Phi ^{' \ze\ze 2} \left[ 1+\frac{2r_H}{r}\Big(\frac{m}{r}-1\Big)\right]\right\}
	 = \int _{r_H} ^{\infty} dr \left[ e^{-\delta}\left(1-\frac{2r_H}{r}\right)\frac{1}{r^2}\frac{Q^2}{f (\Phi)}\right]\ ,
	\end{equation}
provides a constraint independent from the equations of motion that the numerical solutions must obey. This virial identity is also informative in another respect.
	One can show that $1+\frac{2r_H}{r}\Big(\frac{m}{r}-1\Big) >0$. Thus, the left hand side of the equation is strictly positive. Since for $Q=0$ the right hand side would vanish, one concludes that a nontrivial scalar field requires a nonzero electric charge.
	
Second, we can use the linear Smarr-like formula for this family of solutions. This turns out to not have an \textit{explicit} imprint of the scalar field
	\begin{equation}
	M=\frac{1}{2}T_H A_H + \mathcal{V}  Q\ .
	\end{equation}
	One can then compute the first law of BH thermodynamics for EMS BHs, that reads
	\begin{equation}
	dM =\frac{1}{4}T_H dA_H + \mathcal{V} dQ\ .
	\end{equation}
	Third and last, we have noticed that all the solutions also obey a non-linear Smarr formula~\cite{Herdeiro:2018wub},  
	\begin{equation}
	M^2+Q_s^2=Q^2+\frac{1}{4}A^2 _H T_H ^2\  .
	\end{equation}
%
 \section{Subclass IIB solutions: the quartic coupling example}\label{S3}
%
In this paper we are interested in coupling functions $f(\Phi)$ that: $(i)$ allow the existence of both RN and scalarised BHs; $(ii)$ the RN BHs are not unstable against scalar perturbations. From the Klein-Gordon equation
	\begin{equation}
	\Box\Phi =\frac{F_{\mu \nu} F^{\mu \nu}}{4}\dot{f}(0)\ ,
	\end{equation}
the first condition requires $\dot{f}(0)=0$. The second condition, on the other hand, means there should be no tachyonic instablity for RN. A sufficient (but not necessary) condition for this to hold is 
	\begin{equation}
	\ddot{f}(0) =0\ .
	\end{equation}
A possible coupling function that obeys these conditions, and for which $\Phi=0$ is a global minimum, is 
\begin{equation}
f (\Phi ) = 1+\alpha \Phi ^4 \ ,
\label{quartic}
\end{equation}
where $\alpha$ is a dimensionless coupling constant. This defines an example of subclass IIB models. In the following we shall focus on this illustrative example of class IIB.

%
	\subsection{Domain of existence}\label{S32}
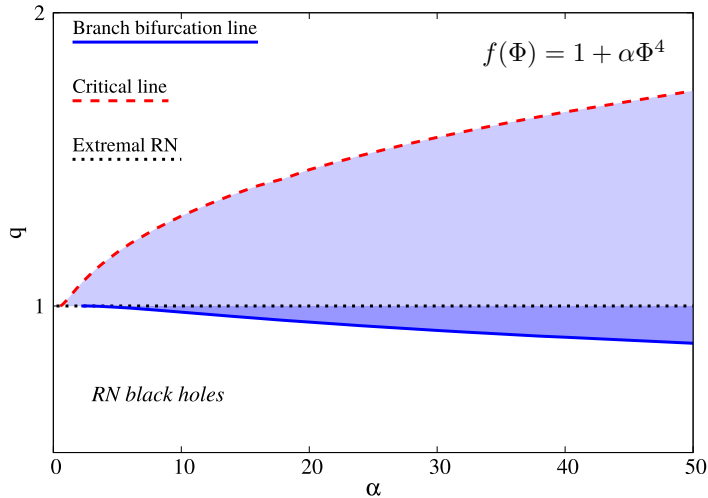
\begin{figure}[h!]
			 \centering
			 \begin{picture}(0,0)
			 \put(185,165){$f (\Phi )=1+\alpha \Phi ^4$}
			\end{picture}
			 \includegraphics[scale=0.75]{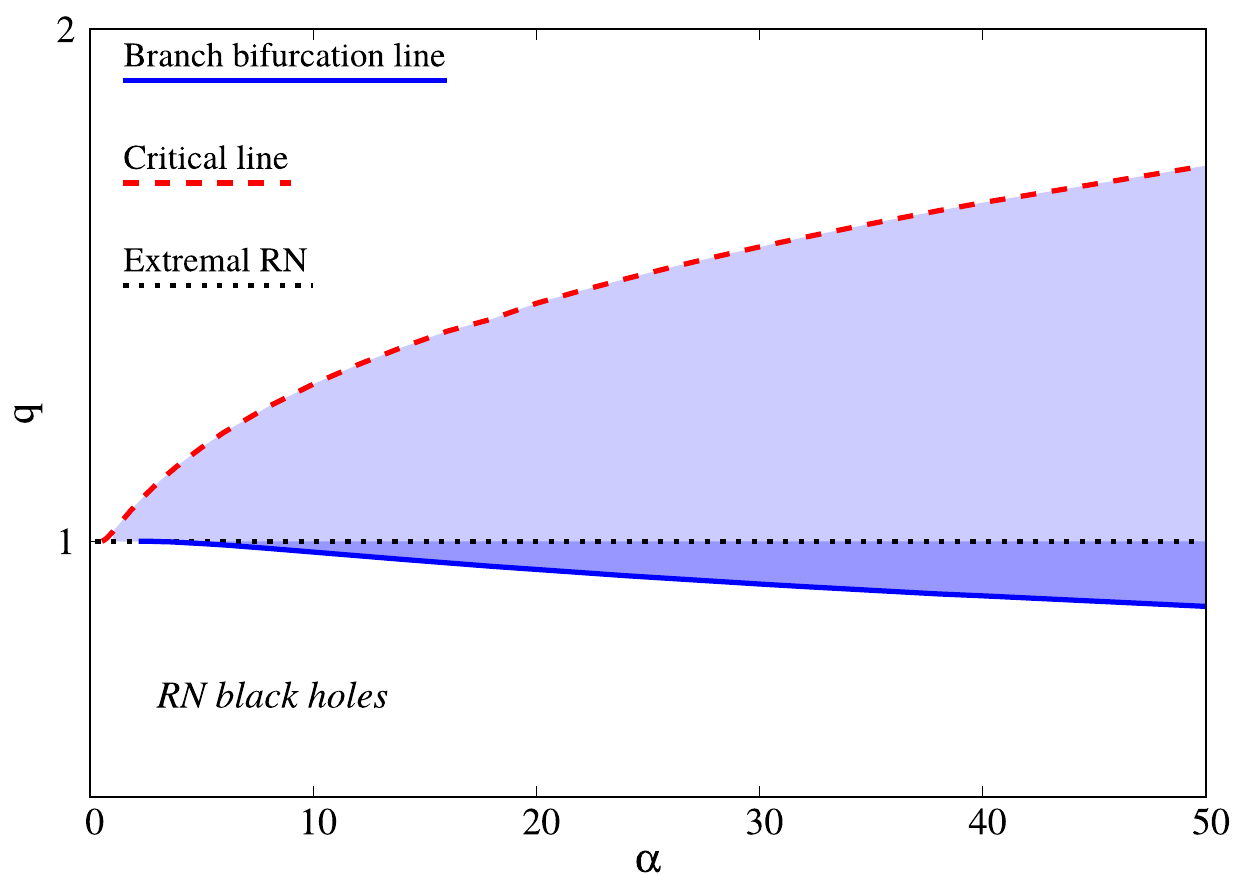}
	 		 \caption{Domain of existence of fundamental BH solutions in EMS models with a quartic coupling (blue shaded region) in a $q$ $vs.$ $\alpha$ diagram. The domain is bounded by the (dashed red) critical line and the (solid blue) branch bifurcation line. In the dark shaded region there are two distinct scalarised BH solutions and a RN solution. In the light shaded region there is only one scalarised BH solution (hot BHs).}
	 		 \label{F1}
	\end{figure}

%
The domain of existence of the fundamental BH solutions ($i.e.$ where the scalar field has no nodes)  of model~\eqref{E21} with coupling~\eqref{quartic}, in a charge to mass ratio $vs.$ coupling constant diagram, is presented in Figure~\ref{F1}.

Fixing a generic value of $\alpha$ one observes the following behaviour.  
A first sequence of (cold) scalarised BHs 
{starts}
from the extremal RN BH. They form a first branch of solutions with  monotonically decreasing $q$ until a minimum value, $q_{\rm min}$, is attained. This minimum value is sub-extremal ($q<1$) and depends on $\alpha$, $q_{\rm min}=q_{\rm min}(\alpha)$. In Figure~\ref{F1} the curve $q_{\rm min}(\alpha)$ is the blue solid branch bifurcation line. At $q_{\rm min}(\alpha)$ a second branch of (hot) scalarised solutions emerges. This branch has monotonically increasing $q$, extending into the over-extremal  regime $(q>1$). The second branch ends at a critical (singular) configuration wherein $a_H=0$, $t_H>0$ and $q=q_{\rm max}>0$. Again, $q_{\rm max}=q_{\rm max}(\alpha)$. The corresponding curve in Figure~\ref{F1} is the red critical line.

\begin{figure}[h!]
			 \centering
	 		 \includegraphics[scale=0.63]{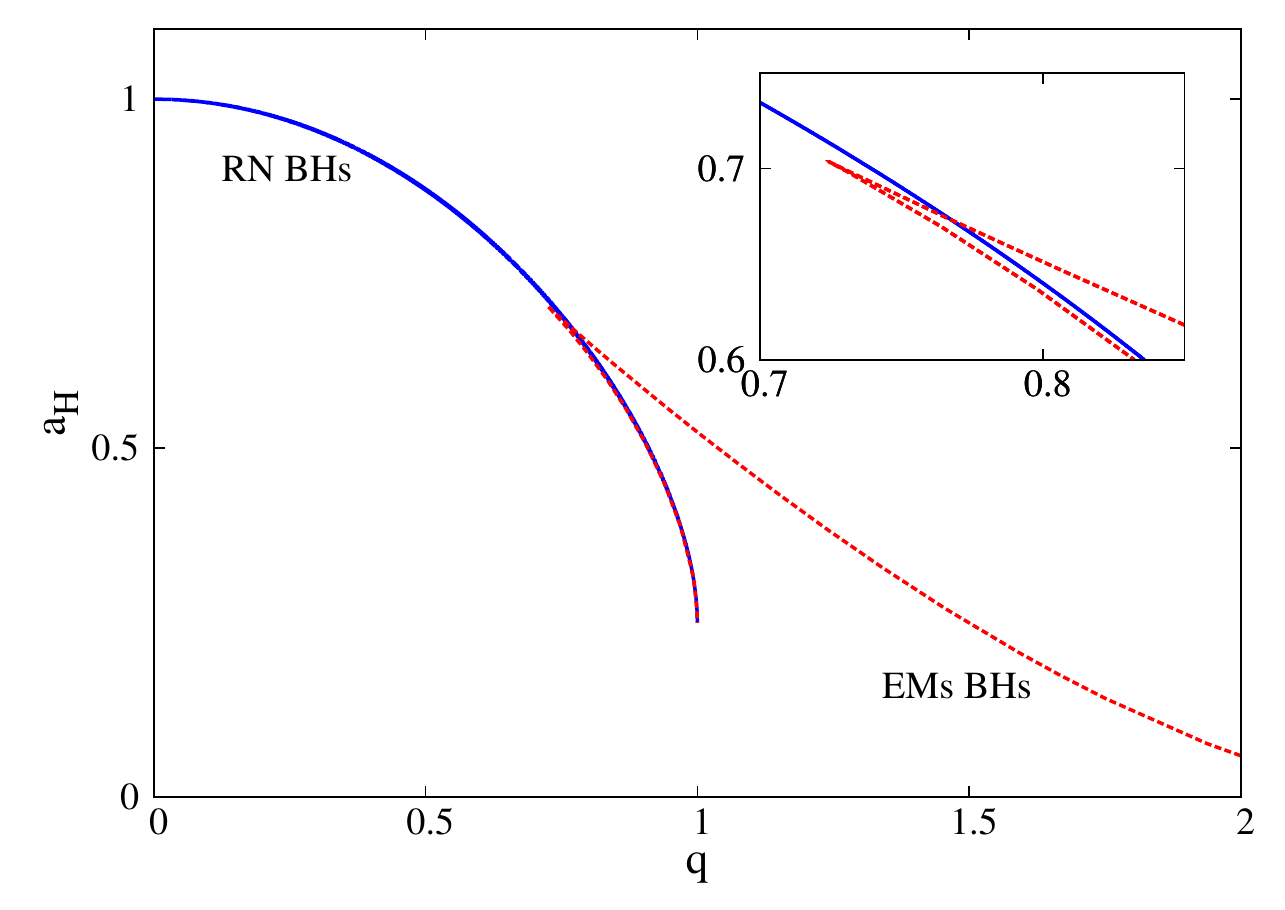}
\includegraphics[scale=0.39]{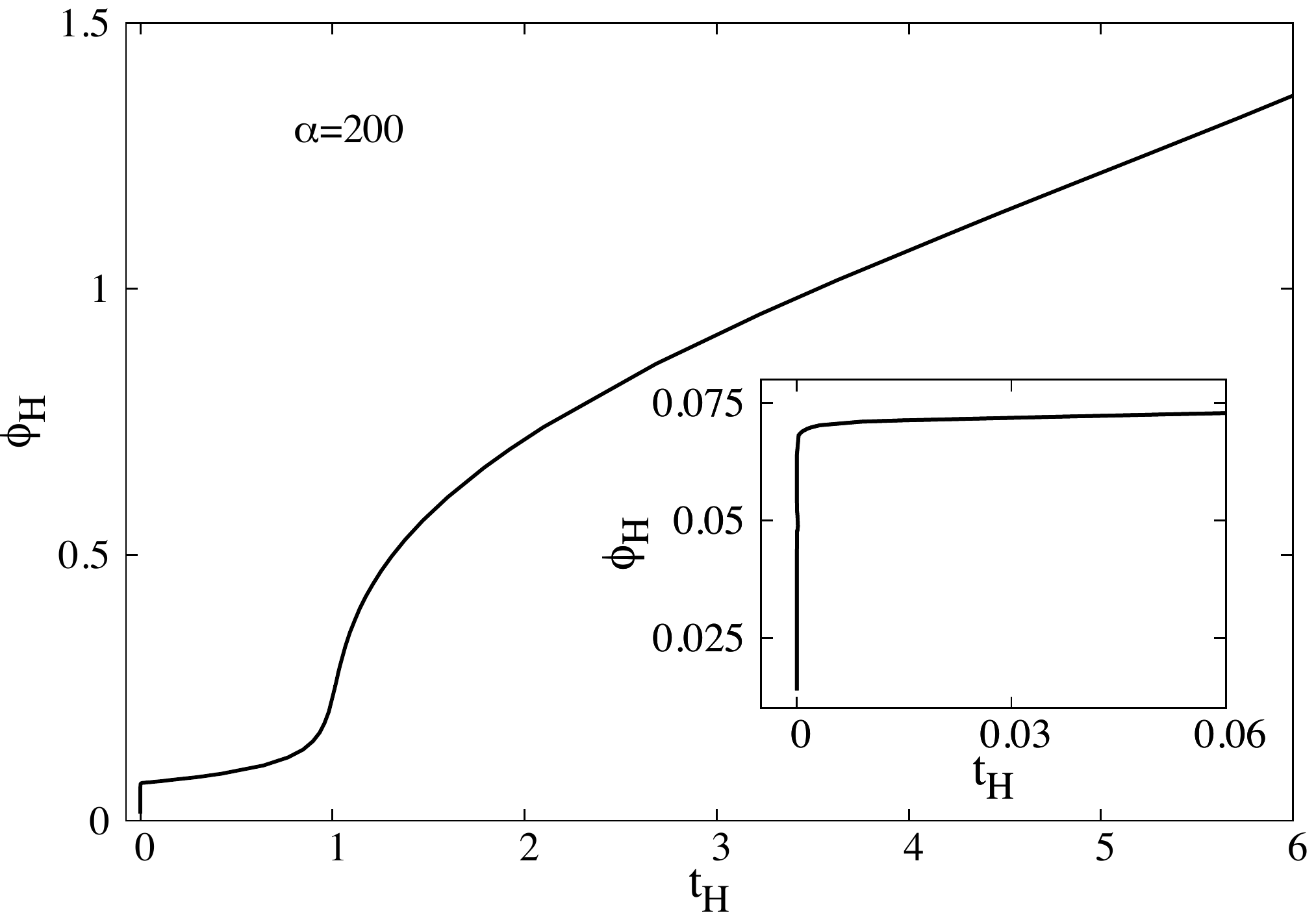}
	 		 \caption{(Left panel) Branches of scalarised BH solutions (red curves, for  $\alpha=20,200$) and RN BHs  (blue curve) in a reduced area $vs.$ reduced charge diagram. (Right panel) Scalar field at the horizon $vs.$ reduced temperature. One may observe the fast supression of $\Phi_H$ as $t_H$ vanishes, showing the scalarised BHs are continuously connected to the extremal RN solution.}
	 		 \label{F2}
	\end{figure}

A complementary perspective on this domain of existence is given in Figure~\ref{F2} (left panel). In this plot one can appreciate the two branch structure of the scalarised BHs. Along the first (cold) branch, emerging from extremal RN, $q$ decreases and $a_H$ increases. Along the second (hot) branch, emerging from the solution with $a_H >0,\ t_H > 0$ and $q=q_{\rm min}$,  $q$ increases and $a_H$ decreases.  Figure~\ref{F2} (left panel) also makes clear the vanishing of the reduced horizon area at the critical solution. The right panel of Figure~\ref{F2}, which shows the value of the scalar field at the horizon, $vs.$ the reduced BH temperature, clarifies that the scalarised solutions in the cold branch continuously connect to the extremal RN, which has $a_H=0.25,\ t_H=0$ and $q=1$.

It is worth emphasising that there is a new sort of non-uniqueness amongst EMS models in this case. In the dark shaded blue region of Figure~\ref{F1}, there are three different solutions for the same $q$, two scalarised ones (cold and hot) and the standard RN. This is qualitatively different from the previous EMS models studied, where at most two solutions with the same $q$ were found in regions of non-uniqueness, corresponding to one scalarised BH and one RN BH, see $e.g.$~ \cite{Herdeiro:2018wub,fernandes2019spontaneous,Astefanesei:2019pfq}. Moreover, whereas in previous IIA models the scalarised BHs were always entropically favoured for the same global charges, this is not the case here, as manifest from Figure~\ref{F2} (left panel). In fact, only in part of the second branch the scalarised BHs have larger area, and hence larger entropy, than the comparable RN BH, $i.e.$ with the same $q$.

\begin{figure}[h!]
			 \centering
			 \includegraphics[scale=0.63]{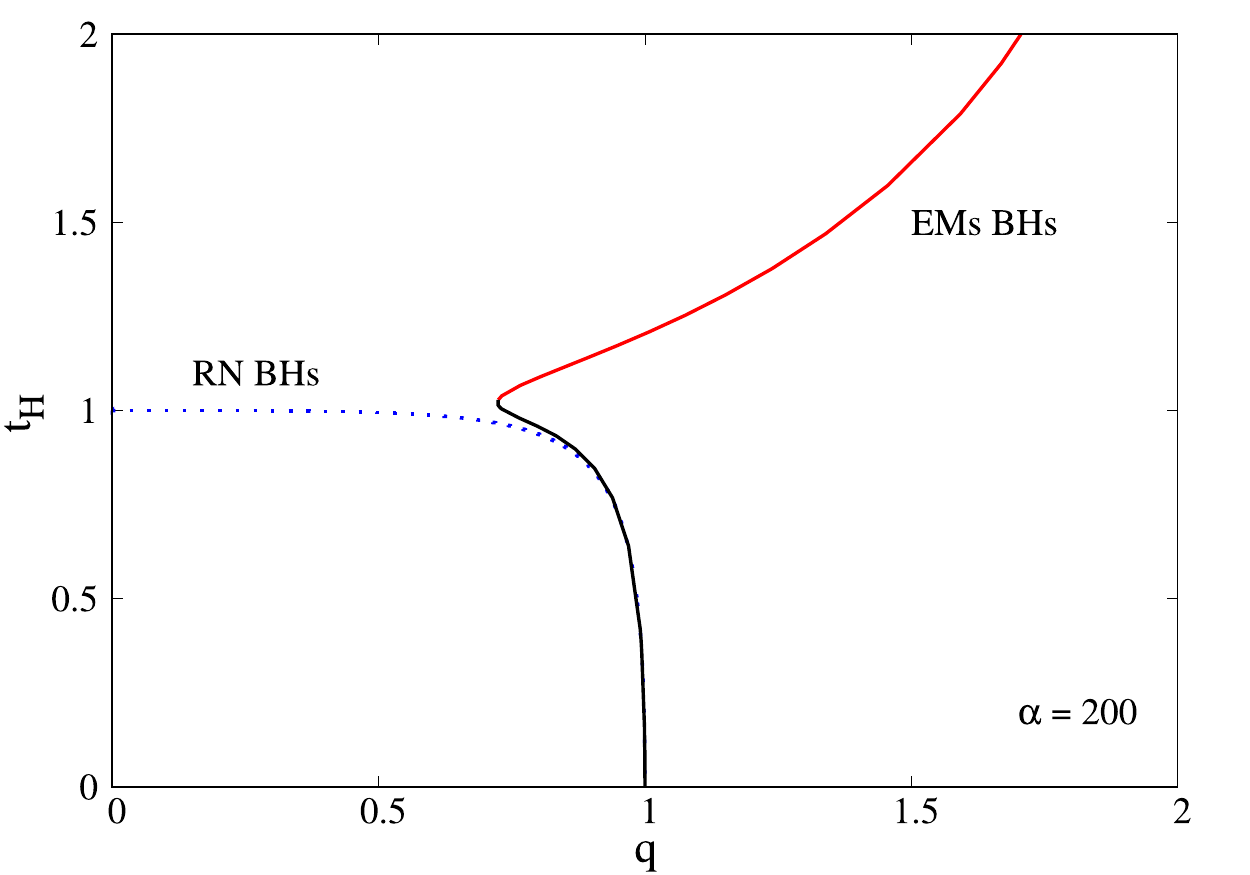}
			
	 		 \caption{Reduced temperature $vs.$ reduced charge for $\alpha=200$. Scalarised BHs are cold in the first branch and hot in the second branch.}
	 		 \label{F3}
	\end{figure}

Finally, Figure~\ref{F3} shows the reduced temperature $vs.$ charge. The first branch of scalarised solutions starts at zero temperature (black solid line). The horizon temperature monotonically increases \textit{both} along the first and second branch (red solid line).  RN BHs correspond to the blue dotted line. They are cooler than the BHs in the second branch. This diagram justifies the terminology cold/hot for the scalarised BHs in the first/second branch.

In Table~\ref{T1} we compare the three degenerate solutions for a specific choice of $\alpha=200$ and $q=0.9$. One can confirm that the scalarised BH in the hot branch has the largest area (and hence entropy), temperature, scalar charge and scalar field value at the horizon. 

        \begin{table}[h!]

                       \centering

                       \caption{Physical properties of three $q$ degenerate BH solutions, for $\alpha =200$ and $q=0.9$.}

                       \vspace{2mm}

                              \begin{tabular}{c|ccccccc}

                                $ $  & $Q_s$ & $\mathcal{V}$ & $a_H$ & $t_H$ & $\Phi _H$\\                    

                               \hline

                         RN         & $0.0000$ & $0.6268$ & $0.5154$ & $0.8457 $ & $0.0000$\\

                         Branch $1$ (Cold) & $0.0378$ & $0.6237$ & $0.5128$ & $0.8555$ & $0.1360$ \\

                         Branch $2$ (Hot) & $0.3934$ & $0.3660$ & $0.5854$ & $1.1454$ &\ $0.4011$


                         \label{T1}

                        \end{tabular}

                   \end{table}

%
\section{Stability}\label{S4}
%
In the model~\eqref{E21} with coupling~\eqref{quartic} the RN BH is stable against all perturbations, as the perturbation equations reduce to the same as in electrovacuum. To consider the perturbative \textit{radial} stability of the scalarised BHs, we take the following ansatz: for the metric 
\begin{eqnarray}
g= 
-N(r)\ze e^{-2  \delta (r)}\Big[1+\epsilon e^{-i\omega t} F_t(r)\Big] dt^2 
+ 
\frac{1}{N(r)}\Big[1+\epsilon e^{-i\omega t} F_r(r)\Big] dr^2
+
r^2 d\Omega^2 \ ;
 \label{eq:metric}
\end{eqnarray}
for the EM field:
\begin{eqnarray}
A= a_0(r)\Big[1+\epsilon e^{-i\omega t} F_0(r) \Big] dt \ ;
\label{eq:EM-field}
\end{eqnarray}
and for the scalar field
\begin{eqnarray}
\Phi= \Phi_0(r) +\epsilon e^{-i\omega t} \Phi_1(r) \ .
\label{eq:scalar-field}
\end{eqnarray}

It is possible to show that, using the field equations, the first order radial perturbations in~\eqref{eq:metric}-\eqref{eq:scalar-field} reduce to a one-dimensional Schr\"odinger-like equation:
\begin{eqnarray}
-\frac{d^2Z}{dR^{*2}} + V_{r}Z = \omega^2 Z \ ,
\label{1ds}
\end{eqnarray}
where $R^*$ is the tortoise coordinate defined by
\begin{eqnarray}
\frac{dR^*}{dr} = \frac{e^{\delta(r)}}{N(r)} \ ,
\end{eqnarray}
the perturbation function is $Z\equiv \Phi_1/r$ and  the potential $V_{r}$ is 
\begin{equation}
V_{r}=\frac{N}{r^2} e^{-2\delta}\left\lbrace 
1-N
-2r^2\Phi_0'^2
+\frac{Q^2}{f}
\left[
2\Phi_0'^2
-\frac{8\alpha}{r f}\Phi_0^3\Phi_0'
-\frac{1}{r^2f^2}
\Big(
\alpha^2 \Phi_0^6(\Phi_0^2-10)
+2\alpha \Phi_0^2(\Phi_0^2+3)
+1
\Big)
\right]
\right\rbrace 
\end{equation}

We are interested in solutions that have an ingoing wave behaviour at the horizon and outgoing at infinity. This fixes the boundary conditions for the perturbation function $Z$. 

Observe that we have fixed the sign convention of the imaginary part of the frequency, $\omega_I$, by choosing the time dependence $e^{-i\omega t}$. Then, unstable modes (that grow in time) have $\omega_I>0$. Indeed,  $e^{-i\omega t}=e^{i\omega_r t} e^{\omega_I t} \to \infty$ as $t\to\infty$, when $\omega_I>0$.
At spatial infinity, $i.e.$ $R^* \to \infty$, we have $Z = A_+ e^{i\omega R^*}$. This means that in the case of an unstable mode of the form $\omega=i\omega_I$, with $\omega_I>0$, $Z = A_+ e^{-\omega_I R^*} \to 0$. 
At the horizon, $i.e.$ $R^* \to -\infty$, we have $Z = A_- e^{-i\omega R^*}$. An unstable mode satisfies $Z = A_- e^{\omega_I R^*} \to 0$. 
Hence unstable perturbations satisfy $Z(r=r_H)=Z(r=\infty)=0$. 

\begin{figure}
	\centering
	\includegraphics[width=0.38\linewidth,angle=-90]{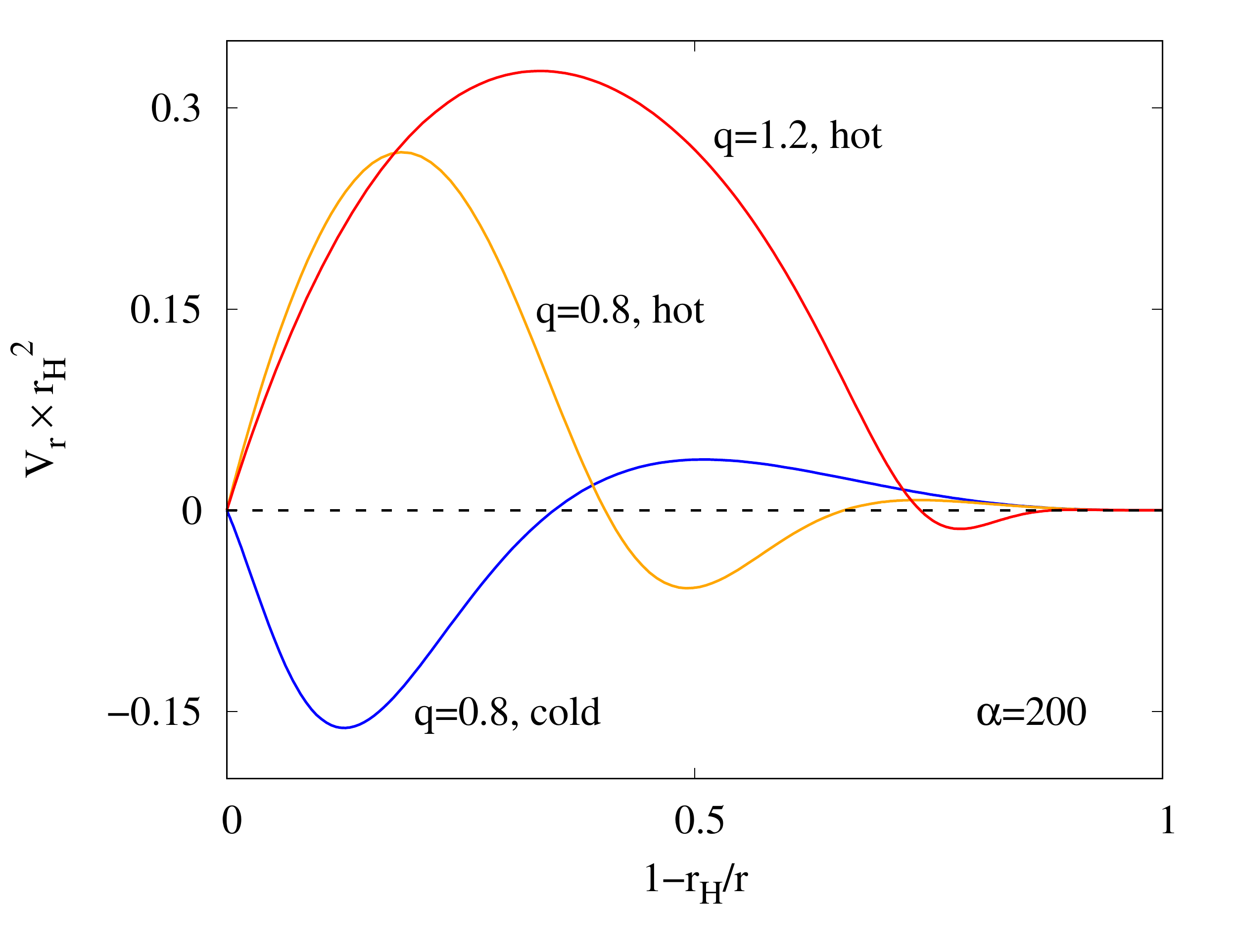}
	\includegraphics[width=0.38\linewidth,angle=-90]{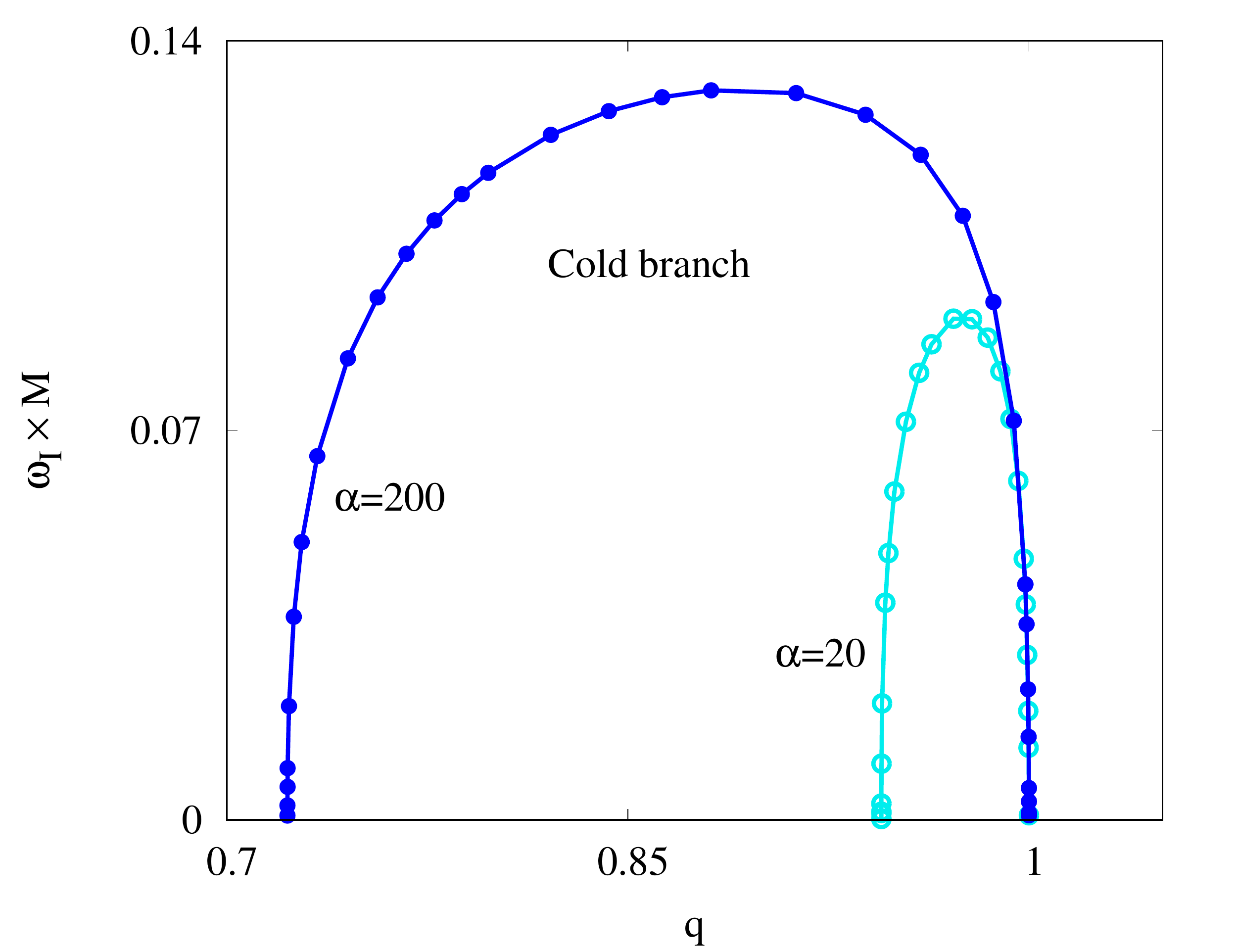}
	\caption{(Left panel) Effective radial perturbations potential as a function of the radial compactified coordinate for one solution in the cold branch and two solutions in the hot branch.  (Right panel) Scaled imaginary part of the mode as a function of $q$ for cold scalarised BHs with $\alpha=20,200$.}
	\label{fig:stability}
\end{figure}

With these boundary conditions, absence of any unstable modes would follow from the positivity of the potential - see $e.g.$~\cite{fernandes2019spontaneous} and below. It turns out, however, that the potential always has a negative region.  For solutions in the cold branch the potential is strongly negative close to the horizon. We show an example in Figure \ref{fig:stability} (left panel), blue curve.
For solutions in the hot branch, on the other hand, the potential also has a negative part; however, this occurs away from the horizon. Moreover, this negative region of the potential becomes smaller along the hot branch, when moving away from the bifurcation point, $i.e.$ for larger $q$. This is illustrated in  Figure \ref{fig:stability} (left panel), orange and red curves. The bottom line of these considerations is that the potential is not positive definite. Consequently, this analysis is inconclusive concerning radial stability.

In order to assess the radial stability of the scalarised solutions we resort to an explicit computation of possible unstable modes.
Following the procedure used in other cases \cite{Blazquez-Salcedo:2018jnn}, we have successfully obtained unstable modes of the previous potential, but \textit{only} for the scalarised solutions in the cold branch. The results are shown in Figure~\ref{fig:stability} (right panel), where we exhibit the positive imaginary part of the mode frequency, scaled by the mass, as a function of $q$. As the figure shows, cold scalarised BHs (first branch) have an unstable mode ($\omega=i\omega_I$, with $\omega_I>0$). The absolute value of $\omega_I$ becomes very small at both end-points of this branch: close to extremality ($q \to 1$) and close to the bifurcation point with the hot branch. For hot scalarised BHs (second branch), however, we could not obtain numerically any unstable mode for any solution in this branch. This includes solutions for which the reduced area is lower than that of a comparable RN BH. Thus, from this analysis we can conclude that cold scalarised BHs are radially unstable, but nothing can be concluded about hot scalarised BHs.

\subsection{S-deformation method}
Since for hot BHs the potential is always negative in some region, unstable modes may exist, albeit we could not find them in the previous section. The $S$-deformation method~\cite{Kimura:2017uor,Kimura:2018eiv,Kimura:2018whv}, however, allows us to show that this is not the case, and that all solutions in the hot branch are in fact radially stable. Let us first briefly explain the procedure and then proof this statement.

Multiplying~\eqref{1ds} by $\bar{Z}$ (where the over bar denotes complex conjugation), integrating from the horizon ($R^*=-\infty$), to $R^*=\infty$, and then using partial integration, we are left with 
\begin{eqnarray}
\int_{-\infty}^{\infty}{ dR^*
	\Big(
	\Big|\frac{dZ}{dR^*}\Big|^2 + V_r |Z|^2
	\Big)
}
= \omega^2 \int_{-\infty}^{\infty}{dR^*|Z|^2} \ ,
\label{identity}
\end{eqnarray}
where we have used the boundary conditions on $Z$ for unstable perturbations.  For unstable modes ($\omega^2<0$), the right side is negative. This means that for unstable modes to exist $V_r$ cannot be strictly positive, a result already quoted above. 

Next, we generalize \eqref{identity} by introducing the $S$-deformation function. To do this, first rewrite~\eqref{1ds} by making use of the identity
\begin{eqnarray}
-\bar{Z}\frac{d^2Z}{dR^{*2}} = 
-\frac{d}{dR^*}\Big(\bar{Z}\frac{dZ}{dR^{*}} \Big)
+ \Big| \frac{dZ}{dR^{*}} \Big|^2 \ .
\end{eqnarray}
Then, introduce an arbitrary $S$ function (deformation function) into the wave equation
\begin{eqnarray}
-\bar{Z}\frac{d^2Z}{dR^{*2}} + V_r |Z|^2 = 
-\frac{d}{dR^*}\Big(\bar{Z}\frac{dZ}{dR^{*}} + S|Z|^2 \Big)
+ \Big| \frac{dZ}{dR^{*}} + SZ\Big|^2 +  \Big(V_r - S^2 + \frac{dS}{dR^*}\Big) |Z|^2  
= \omega^2 |Z|^2 \ .
\end{eqnarray}
Now repeat the integration. After using partial integration we get 
\begin{eqnarray}
-\Big[\bar{Z}\frac{dZ}{dR^*} + S|Z|^2\Big]_{-\infty}^{\infty}
+\int_{-\infty}^{\infty}{ dR^*
	\left[
	\Big|\frac{dZ}{dR^*} + SZ\Big|^2 + \left(V_r - S^2 + \frac{dS}{dR^*} \right)|Z|^2
	\right]
}
= \omega^2 \int_{-\infty}^{\infty}{dR^*|Z|^2} \ .
\end{eqnarray}

Next, we restrict the possible $S$ functions. We assume this function is smooth everywhere and it does not diverge at the boundaries. These conditions, together with the boundary condition for unstable perturbations make the first term of the previous expression vanish. We are left with
\begin{eqnarray}
\int_{-\infty}^{\infty}{ dR^*
	\Big|\frac{dZ}{dR^*} + SZ\Big|^2
}
+
\int_{-\infty}^{\infty}{ dR^*
	 \Big(V_r - S^2 + \frac{dS}{dR^*} \Big)|Z|^2
}
= \omega^2 \int_{-\infty}^{\infty}{dR^*|Z|^2} \ .
\end{eqnarray}
The first term in the left hand side is positive. The right hand side is negative, if $\omega^2<0$. Thus, if we show that the second term of the left hand side is positive or vanishes, we establish no modes with $\omega^2<0$ are possible, for this potential. Observe that this does not require the potential $V_r$ to be strictly positive anymore.

In practice, the absence of unstable modes is established by defining the deformed potential $\hat{V}_r$~\cite{Kimura:2017uor,Kimura:2018eiv,Kimura:2018whv}:
\begin{eqnarray}
\hat{V}_r=V_r + \frac{dS}{dR^*} - S^2 \ .
\end{eqnarray}
Then, it is enough to show that it is possible to make $\hat{V}_r=0$. This implies the original potential does not contain unstable modes. Such condition defines a Riccati-type differential equation 
\begin{eqnarray}
\frac{dS}{dR^*}=S^2-V_r \ ,
\end{eqnarray}
or
\begin{eqnarray}
\frac{dS}{dy}=\frac{dr}{dy}\frac{dR^*}{dr}(S^2-V_r) \ ,
\label{eq:S}
\end{eqnarray}
where $y=1-r_H/r$, $dr/dy=r_H/(1-y)^2$ and $dR^*/dr=e^{\delta}/N$. Since the potential is zero at $r=r_H$ ($y=0$) and $r=\infty$ ($y=1$), a solution of (\ref{eq:S}) has to satisfy $S(y=0)=S(y=1)=0$.

\begin{figure}
	\centering
	\includegraphics[width=0.38\linewidth,angle=-90]{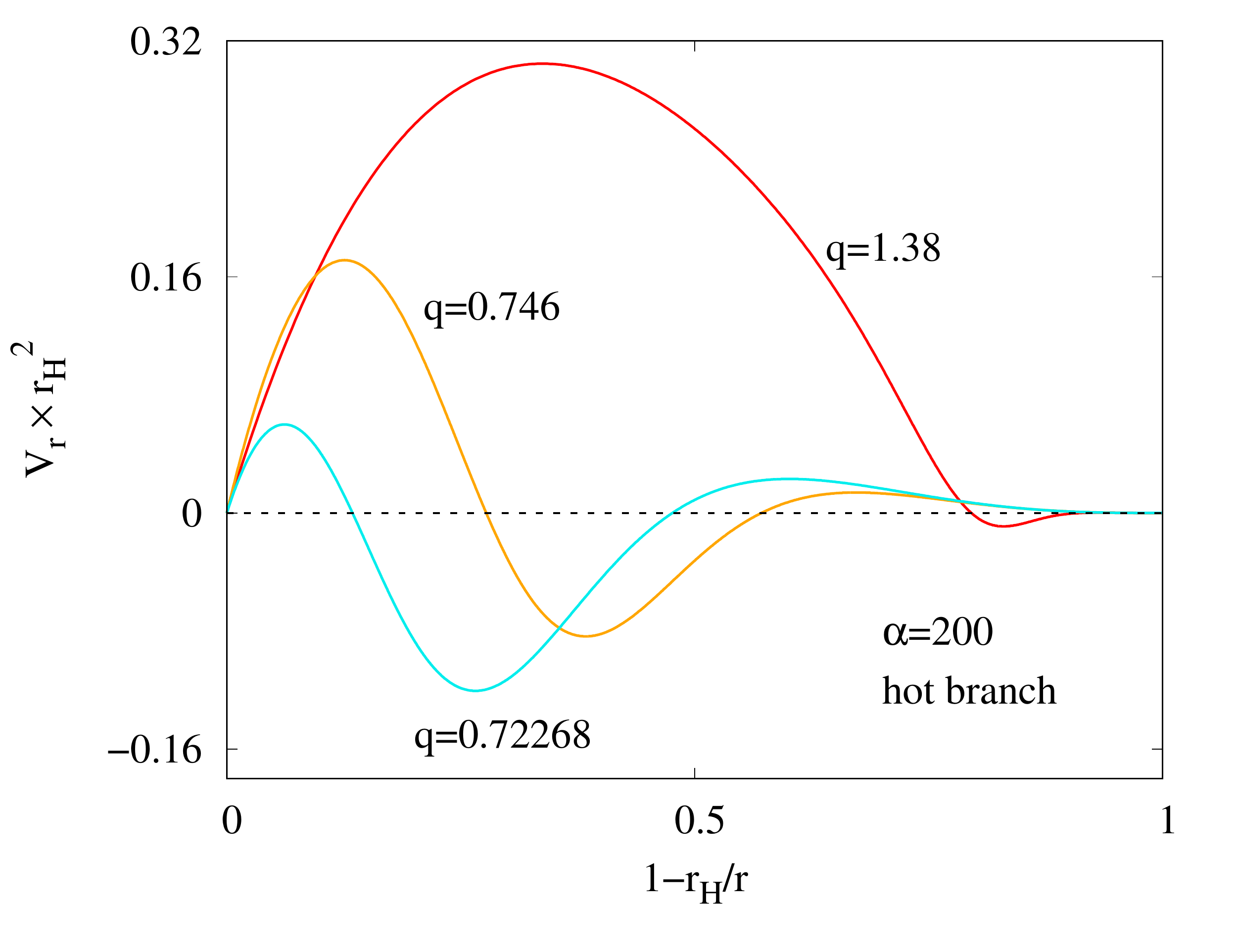}
	\includegraphics[width=0.38\linewidth,angle=-90]{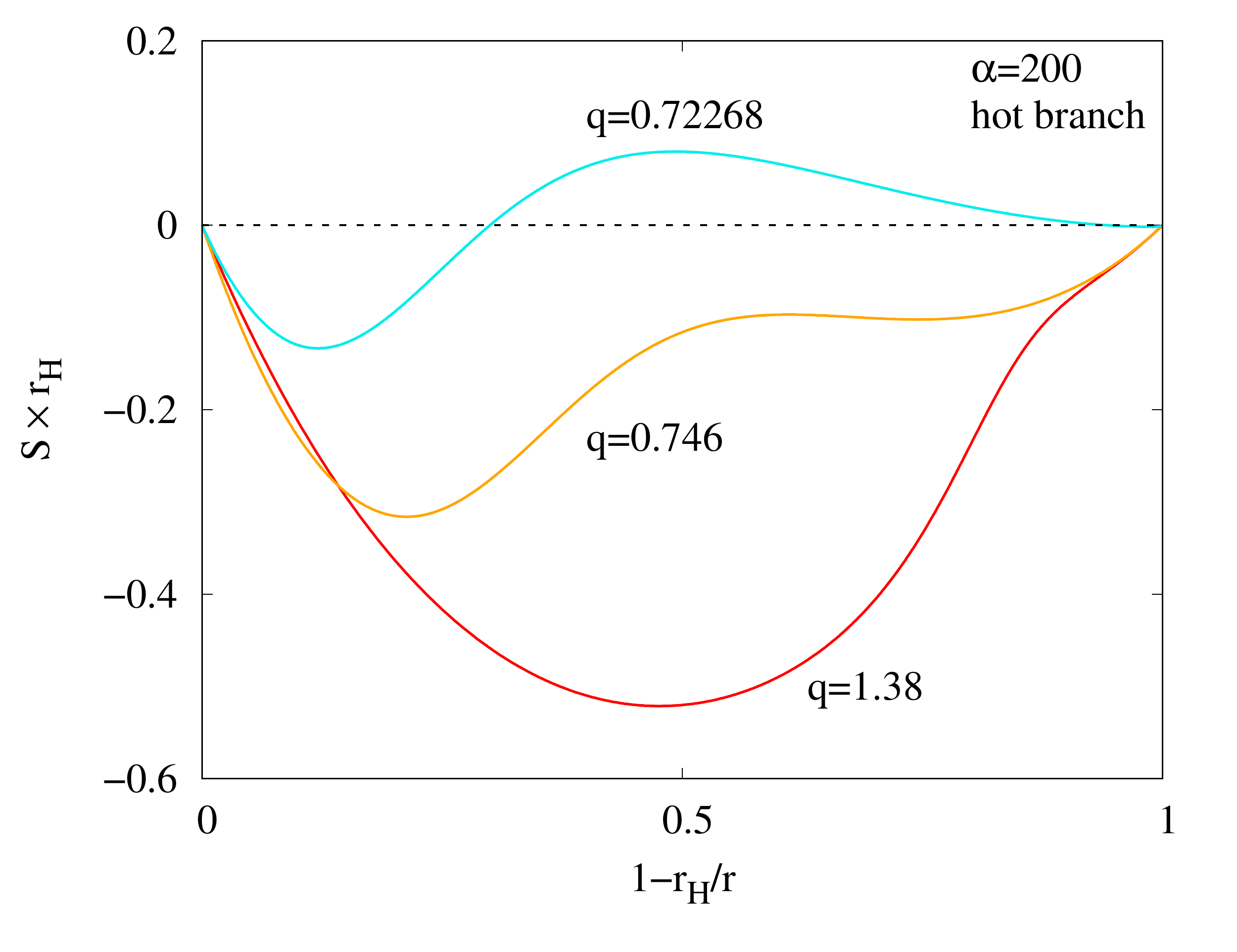}
	\caption{(Left panel) Effective radial potential as a function of the radial compactified coordinate for several hot BH solutions. (Right panel) $S$-deformation function for the same solutions.}
	\label{fig:Sdeformation}
\end{figure}

We have numerically integrated (\ref{eq:S}), reading off the potential $V_r$ from the numerical scalarised BH solutions.  A few examples of the potential for hot scalarised BHs (interpolating between points with a cubic spline) are shown in Figure~\ref{fig:Sdeformation} (left panel). The differential equation is solved with boundary condition $S(0)=0$ in a domain $y\in[0,1]$. The result of the integration is that we were able to find solutions for $S(y)$ - Figure~\ref{fig:Sdeformation} (right panel). The deformation function approaches zero at the right side. It was possible to obtain the deformation function for all hot BH solutions we have tackled. This includes solutions for which the reduced area is lower than that of a comparable RN BH - $e.g.$ the blue curve in Figure~\ref{fig:Sdeformation}.  Thus, since we have managed to obtain a regular $S$ function for the BH solutions in the second branch, hot scalarised BHs are radially stable, even though the original potential is not strictly positive everywhere.

As a final remark, if the same procedure is attempted for solutions in the first branch, one cannot integrate (\ref{eq:S}); the equation develops a singularity. This is consistent, of course, with the fact that we have numerically obtained unstable modes for cold BHs. Thus, it is clear that the $S$ function cannot exist.

%
\section{Remarks and a curious analogy}\label{S5}
%
In this work we have analysed EMS models of subclass IIB, according to the classification in~\cite{Astefanesei:2019pfq}. In these models there are both RN and scalarised BHs, but the RN BHs do not suffer any perturbative instability triggered by the scalar field. Our analysis focused on the particular case of a quartic coupling function and unveiled a number of qualitatively distinct features, as compared to the previously analysed EMS models of class IIA.

Amongst the novel features, we have unveiled a new type of non-uniqueness, amongst EMS models. For some regions of the domain of existence there are three different BH solutions, the scalar-free (or bald) RN BH and the scalarised (or hairy) cold and hot BHs. As described in the introduction this is qualitatively different from what occurs in subclass IIA EMS models; but it exhibits a very curious analogy with a model which, \textit{a priori}, seems completely unrelated. This analogy pertains to five dimensional vacuum Einstein gravity.  

An influential result in BH physics was the discovery of the vacuum black ring in five dimensional Einstein's gravity~\cite{Emparan:2001wn,Emparan:2001wk}. Black rings come in two types (fat and thin) and co-exist with the Myers-Perry BH~\cite{Myers:1986un} in five dimensional vacuum gravity, being distinguished by their event horizon topology. 
They carry two physical parameters, mass $M$ and angular momentum $J$, but they are typically characterised by a set of reduced quantities: respectively, the reduced angular momentum, horizon area and temperature:
\begin{equation}\label{E320}
 j=\frac{3}{4}\sqrt{\frac{3 \pi}{2 }}\frac{J}{M^{3/2}}\ ,\qquad \quad
a_H=\frac{3}{16} \sqrt{\frac{3}{\pi}} \frac{A_H}{M^{3/2}}\ ,\qquad \quad
t_H=4 \sqrt{\frac{\pi}{3}} T_H \sqrt{M}\ . 
\end{equation}
The overall factors in the above expressions are taken to agree with the usual conventions in the literature~\cite{Emparan:2006mm,Emparan:2007wm}. 

\begin{figure}[h!]
			 \centering
	 		 \includegraphics[scale=0.63]{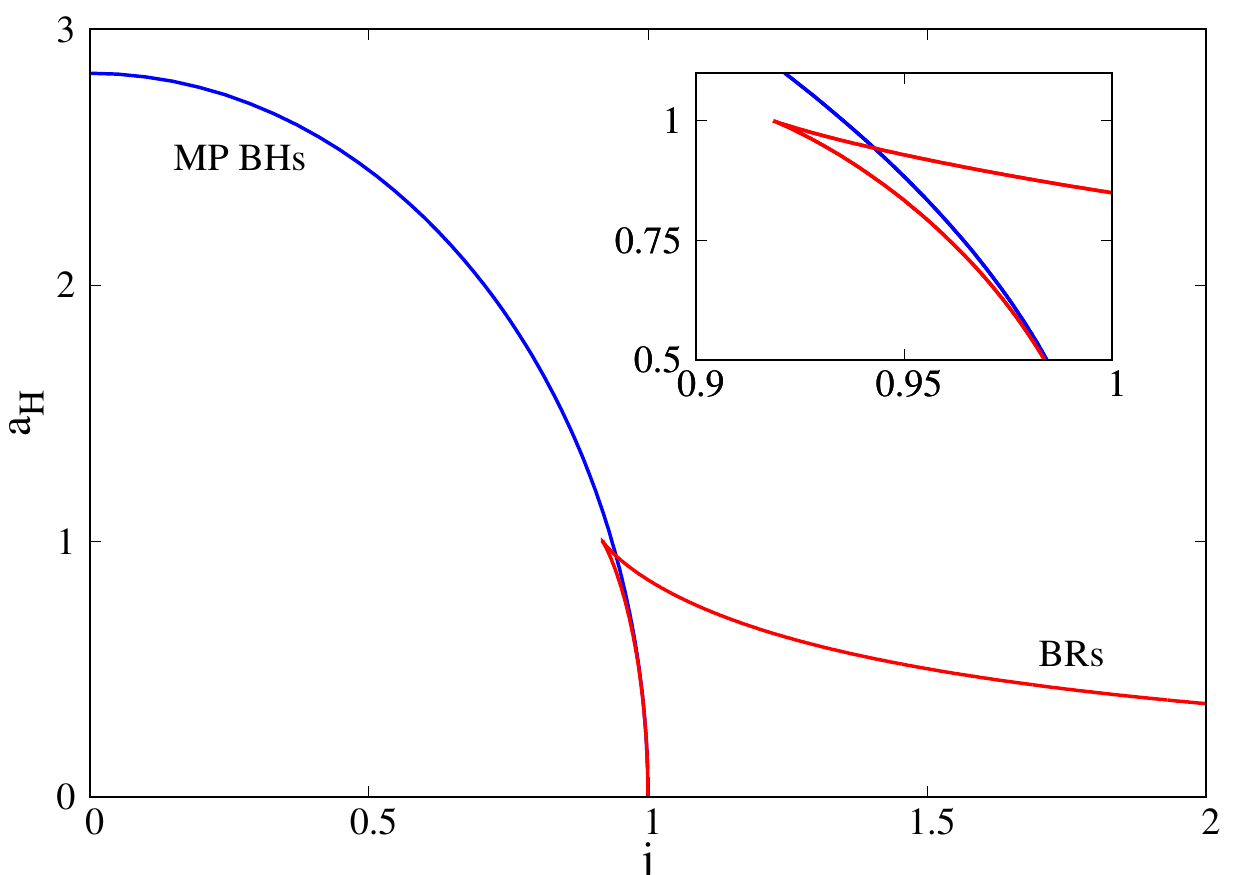}
 \includegraphics[scale=0.63]{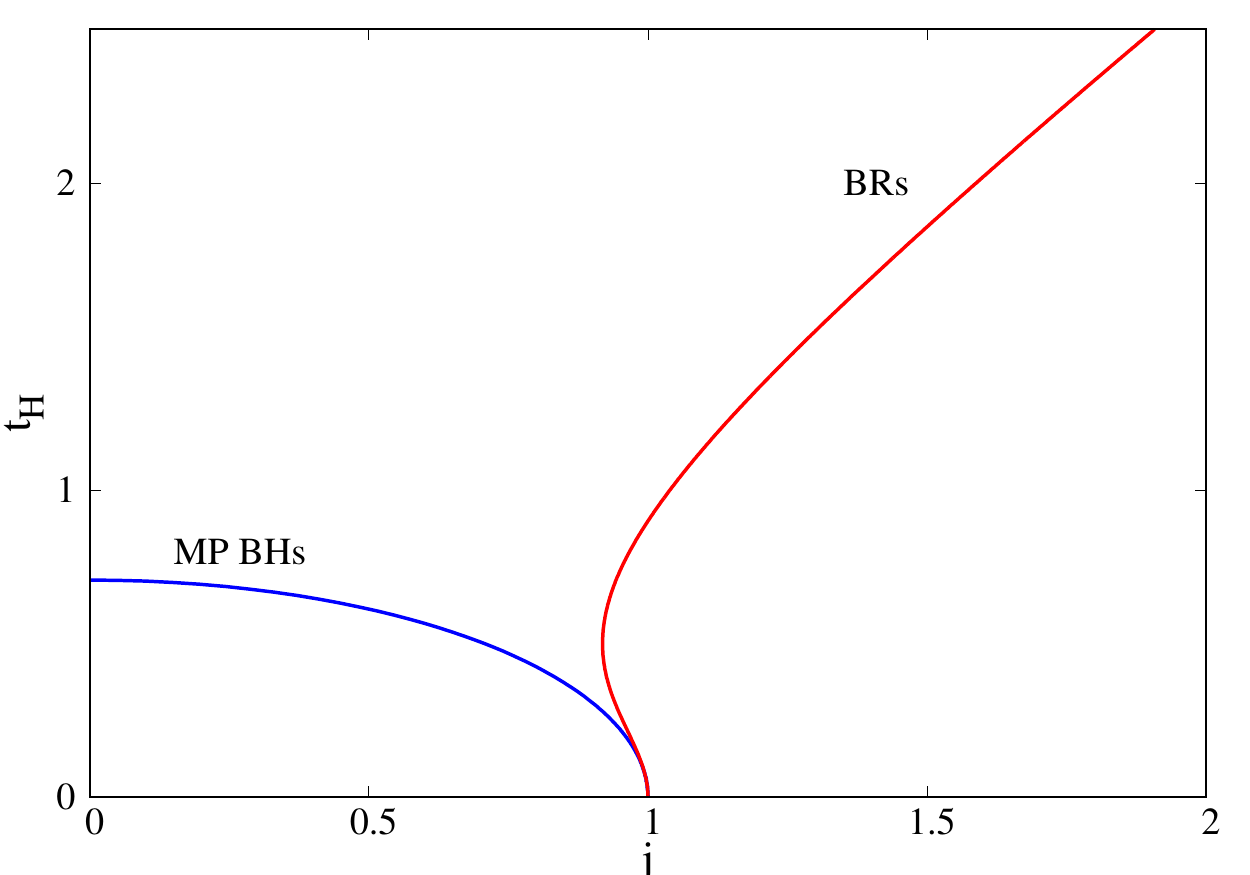}
	 		 \caption{(Left panel) Branches of five dimensional vacuum gravity black rings (red curves) and Myers-Perry BHs  (blue curve) in a reduced area $vs.$ reduced angular momentum diagram. (Right panel) Reduced temperature $vs.$ reduced angular momentum for the same solutions. Black rings are colder when they are fat (in the first branch) and hotter when they are thin (in the second branch).}
	 		 \label{F22}
	\end{figure}

In Figure~\ref{F22} we exhibit the BHs of vacuum five dimensional gravity in a reduced area (left panel) and a reduced temperature (right panel) $vs.$ reduced angular momentum plot.
The parallelism with Figure~\ref{F2} (left panel) and~\ref{F3} is uncanny, with [Myer-Perry BHs, fat rings, thin rings] playing the role of [RN, cold scalarised, hot scalarised] BHs and the reduced angular momentum being mapped to the reduced charge.  

In particular, one observes that the Myers-Perry (RN) BHs exist for a finite range of $0\leqslant j \leqslant 1$ ( $0\leqslant q \leqslant 1$),  where the solution with $j=0$ ($q=0$) corresponds to the Tangherlini (Schwarzschild) BH. Black rings (scalarised BHs), on the other hand, can be over-rotating (over-charged).
At $j=1$ the fat black rings and the Myers-Perry  BHs
 degenerate
to the same  
 (singular) extremal solution. 
 On the other hand, the cold scalarised BHs connect with the (regular) extremal RN BH at $q=1$. 
Fat black rings become thinner, with lower $j$ and larger $a_H$ until a bifurcation point, where they become thin black rings. Cold scalarised BHs become hotter, with lower $q$ and larger $a_H$ until a bifurcation point, where they become hot scalarised BHs. Thin black rings (hot scalarised BHs) become over-spinning (over-charged). At the moment, however, no further weight can be given to this curious analogy.

Let us close with two questions and one remark. Firstly, how generic is the behaviour of the quartic coupling within class IIB? Secondly, are the hot scalarised BHs stable against all perturbations? If so, this model would offer the remarkable situation of two stable BH local ground states. Or, alternatively, are these spherical BHs unstable against non-spherical perturbations?  Either possibility provides an exceptional scenario in BH physics and answering this question is certainly a meritable research programme. Finally, all solutions herein are fundamental states, in the sense the scalar field has no nodes. We have checked that excited solutions also exist, with their own hot and cold branches. However, our results indicate that all of these excited solutions are unstable under radial perturbations.

\medskip
%
\section*{Acknowledgements}
%
A. Pombo is supported by the FCT grant PD/BD/142842/2018. This work is supported by the Center for Research and Development
in Mathematics and Applications (CIDMA) through the Portuguese
Foundation for Science and Technology 
(FCT - Funda\c c\~ao para a Ci\^encia e a Tecnologia),
references UIDB/04106/2020 and UIDP/04106/2020 and by national funds (OE), through FCT, I.P., in the scope of the framework contract foreseen in the numbers 4, 5 and 6 of the article 23, of the Decree-Law 57/2016, of August 29,
changed by Law 57/2017, of July 19. We acknowledge support  from the projects PTDC/FIS-OUT/28407/2017 and CERN/FIS-PAR/0027/2019.   This work has further been supported by  the  European  Union's  Horizon  2020  research  and  innovation  (RISE) programme H2020-MSCA-RISE-2017 Grant No.~FunFiCO-777740.
JK and JLBS gratefully acknowledge support by the DFG funded
Research Training Group 1620 ``Models of Gravity''.
JLBS would like to acknowledge support from the DFG project BL 1553. The authors would like to acknowledge networking support by the
COST Actions CA16104 and CA15117. 


  \bibliographystyle{ieeetr}
  \bibliography{biblio}


\end{document}